\documentclass{ws-procs975x65}

\begin{document}

\newcommand{\tc}{T$_{\rm c}$}

\title{Experimentally Realized Correlated Electron Materials:  From Superconductors to Topological Insulators}

\author{C. KALLIN}

\address{Department of Physics \& Astronomy, McMaster University,\\
Hamilton, Ontario L8S 4M1, Canada\\
$^*$E-mail: kallin@mcmaster.ca}

\begin{abstract}
Recent discoveries, as well as open questions, in experimentally realized correlated electron materials are reviewed.  In particular, high temperature superconductivity in the cuprates and in the recently discovered iron pnictides, possible chiral p-wave superconductivity in strontium ruthenate, the search for quantum spin liquid behavior in real materials, and new experimental discoveries in topological insulators are discussed.
\end{abstract}

\keywords{Superconductors; Frustrated magnets; Topological insulators.}

\bodymatter

\section{Introduction}\label{intro:sec1}
Nature has provided us with an incredibly diverse variety of materials which exhibit striking phenomena driven by electron correlations.  Partially driven by the discovery of new materials, there has been significant recent progress in understanding correlated electron systems which do not fit into our wide-reaching paradigms of Fermi liquid theory and spontaneous symmetry breaking order, although many challenging open questions remain.  It is not possible to review all the interesting strongly correlated materials which are currently under active investigation, so I will primarily focus on some of the newer superconductors, as these are not covered elsewhere in the Proceedings and these especially have stimulated enormous scientific effort, leading to many new ideas.  In particular, I will focus on the high temperature cuprate superconductors, the iron pnictides, and possible chiral p-wave superconductivity in strontium ruthenate.  Newly discovered topological insulators will also be discussed.  A number of related experimental systems are discussed elsewhere in the Proceedings.  In particular, frustrated quantum magnets (mentioned briefly below) are treated in depth by Sachdev\cite{sachdev-solvay}, and quantum Hall systems are the subject of the article by Stern\cite{stern-solvay}.

\section{Unconventional Superconductors}

Unconventional superconductors are those in which superconductivity arises from direct electron-electron interactions, as contrasted to the conventional indirect interaction via phonons.  Direct interactions often favor higher (than s-wave) angular momentum pairing.  Although the normal state of the high temperature superconducting cuprates is not a conventional Fermi liquid, so the concept of pairing electron-like quasiparticles may not be completely valid, it is known that the on-site Coulomb repulsion and spin fluctuations play a key role in stabilizing the d-wave superconducting state.  A completely new class of high temperature superconductors, the iron pnictides, were discovered just in the last year  and their pairing symmetry is still under investigation, as is the question of whether the mechanism for superconductivity in these Fe-based materials is closely connected to that of the cuprates or whether a new route to high temperature superconductivity has been found.  Another novel superconductor which has attracted considerable recent attention is strontium ruthenate, Sr$_2$RuO$_4$.  Ferromagnetic spin fluctuations are believed to be responsible for the superconductivity in this material, but the interest here is not due to a high transition temperature (in fact, \tc\ is only 1.5K) but because experiments point to a chiral p-wave order, which is a topological order that can, under certain conditions, support quasiparticles with non-Abelian statistics.  Intense effort in understanding each of these novel superconductors has led to many new ideas and new paradigms about the type of behavior quantum many-body systems can exhibit.  Specific highlights in our current understanding as well as open questions surrounding each of these superconductors are reviewed below.

\subsection{High Temperature Cuprate Superconductors}

Over the last two decades, the superconducting cuprates have been the most intensely studied materials in physics.  Much of this interest stems from the high superconducting transition temperatures, \tc , and the consequent potential for new applications.  Whereas the maximum observed \tc\ had slowly increased from 4.2K in 1911 (in Hg) to 23K in 1974 (in Nb$_3$Ge), following the discovery in 1986 of superconductivity at 35K in La$_{2-x}$Ba$_x$CuO$_4$,\cite{bednorz-muller} the highest \tc\ quickly shot up to 138K (or higher under pressure) as many other cuprate oxides were discovered.\cite{hg}

Intense interest in the cuprates also follows from the strong role that  electron-electron interactions play in these materials.  Although the correct and complete theory of high temperature superconductivity is still under debate, much is now understood about the behavior of these materials.  More generally, attempts to understand strong electronic correlations in the cuprates have generated many new ideas, particularly in the area of quantum magnetism, as discussed by Sachdev in these Proceedings.\cite{sachdev-solvay}  Research in the cuprates has led to a much deeper understanding of non-Fermi liquid behavior, particularly quantum order or topological order.\cite{wen}

\begin{figure}[t]
\begin{center}
\psfig{file=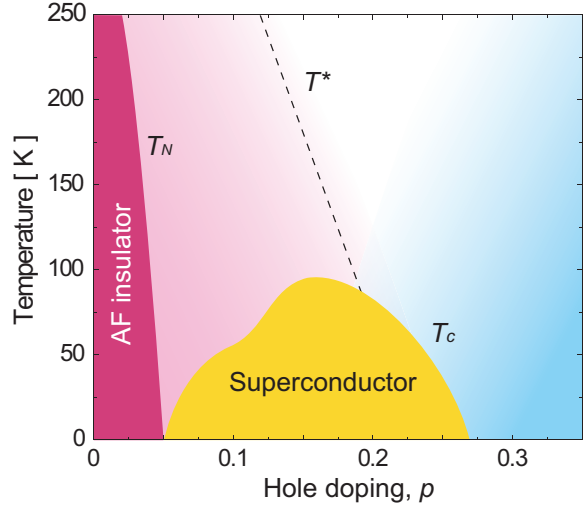,width=3in}
\end{center}
\caption{Cuprate phase diagram as a function of hole doping.  The doping where the maximum in \tc\ is achieved is referred to as optimal doping. The pseudogap phase appears below the crossover temperature, T*.  From Ref.~\refcite{louis}.}
\label{ck:fig1}
\end{figure}

Many different materials belong to the class of cuprate superconductors.  A few well-studied examples are La$_{2-x}$Sr$_x$CuO$_4$, YBa$_2$Cu$_3$O$_{6+x}$ and BiSr$_2$CaCu$_2$O$_{6+x}$. What the cuprates all have in common is fairly weakly coupled copper oxide layers (called planes) which are where all the electronic action is.  The material between these planes acts as a charge reservoir, and changing the crystal stoichiometry ({\it i.e.} changing $x$\ in the chemical formula) changes the electron density, or the ``doping'', p, of the copper oxide layers.  This leads to a temperature versus doping phase diagram, as shown in Fig.~\ref{ck:fig1}.

The undoped phase corresponds to exactly one electron per Cu site, which band theory would predict to be a metal.  However, the undoped phase is a Mott insulator, due to electron-electron interactions.  A strong on-site Coulomb potential, U, localizes the electrons, one to each Cu site.  The electronic correlations also lead to antiferromagnetic order in this phase.  With increasing doping, the material becomes a superconductor. The pairing order parameter is known to have d-wave symmetry, with nodes in the superconducting gap along the directions $k_x=\pm k_y$.\cite{vanharlingen,tsuei} While weak-coupling BCS theory for a d-wave order parameter can be a useful starting point for describing the superconducting phase at low temperatures, there are important deviations from BCS theory.  In particular, in the underdoped region (doping less than the optimal doping where the maximum \tc\ is achieved), the superfluid density and \tc\ fall off with decreasing doping, whereas the superconducting gap increases.\cite{sutherland}  This behavior is believed to result from strong correlation physics and to be a signature of a doped Mott insulator.  Indeed, it follows quite naturally from strong correlation theories inspired by Anderson's resonating valence bond picture.\cite{palee}  At sufficiently large doping, the normal state appears to be a more or less conventional Fermi liquid, whereas in the underdoped and optimally doped region, the normal state is anomalous.  Shown in Fig.~\ref{ck:fig1} is a cross-over temperature, T*.  The anomalous state below T* is called the pseudogap phase, since the low-energy density of states and the spin susceptibility are suppressed in this phase.  There is no observed phase transition at T*, but most physical properties undergo a smooth but substantial change at this cross-over temperature.

The key theoretical goal underlying research in high temperature superconductivity is to explain all universal properties in the insulating, pseudogap and superconducting phases within a theory which can make verifiable predictions.  Much of the current attention is focussed on the pseudogap region for this purpose.  Part of the reason is that, while the ground states of the Mott insulating and the d-wave superconducting phases are understood, the nature of the pseudogap ground state, or whether it is even connected to a ground state as opposed to being a strongly fluctuating phase associated with the insulating and superconducting phases nearby, is still a point of debate.\cite{pseudogs} Furthermore, the pseudogap phase is generally viewed as the key to understanding the cuprates since it occupies a large region of the phase diagram in temperature and doping, it connects the strongly correlated Mott insulating phase to the high temperature superconducting phase, and, most importantly, it is the normal phase from which the superconductor condenses over much of the superconducting dome.  Understanding the pseudogap phase is seen as equivalent to understanding the doped Mott insulator.

There are many different ideas and proposals for the pseudogap phase, including preformed Cooper pairs,\cite{preformed} antiferromagnetic and/or superconducting fluctuations,\cite{so5,affluc} static and fluctuating stripes or nematic order,\cite{stripes} staggered flux\cite{piflux} and d-density wave\cite{ddens} phases, and orbital currents.\cite{orbcur} The staggered flux phase emerges from the resonating valence bond (RVB) picture, which captures much of the cuprate phenomenology.\cite{rvb}  Part of the difficulty in understanding the pseudogap phase is that experiments see evidence for many of these different behaviors, at least in some materials in some parts of the pseudogap region, and it is then a question of which one, if any, is key to high temperature superconductivity.  Below, in these proceedings, Varma\cite{solvay-varm} makes the case for the importance of orbital currents, as several experiments have seen evidence for this order in the pseudogap phase.\cite{mook}  Kivelson and others\cite{stripes} have argued that fluctuating stripes may be central to high \tc\ superconductivity. Fluctuating stripes, as well as the preformed pairs proposal, might be considered precursor theories, in that the proposed order is a precursor to obtaining high temperature superconductivity.  Other proposals can be classified as competing orders, order which competes with superconductivity and leads to the distinctive phase diagram observed. D-density waves\cite{ddens} are an example of competing order, as are static stripes.  Detailed studies of the phase diagram may distinguish between precursor and competing theories as one would expect the crossover temperature, T*, to slice through the superconducting dome, presumably ending in a quantum critical point at T=0 under the superconducting dome in the case of competing order.  By contrast, one would expect T* to hug the superconducting dome, merging together with \tc\ on the overdoped side if the pseudogap is a precursor effect.  In fact, both types of behaviour have been seen in experiment, depending on which physical property or signature one tracks at T*, suggesting that both precursor and competing signatures are present in the pseudogap phase.\cite{pseudophase} In addition to T*, there is a lower cross-over temperature below which one observes an unusual Nernst signal, which is interpreted as evidence for superconducting pairing without long-range phase coherence.\cite{nernst}

Here, I will focus on one particular set of experiments which address the nature of the pseudogap phase and which have generated enormous interest -- recent observations of quantum oscillations in the pseudogap region.\cite{louis}  First, let me briefly review the relevant ARPES results.  In the overdoped regime, a single large Fermi surface, centered at ($\pi,\pi$) and enclosing 1+p holes per Cu site, where p is the hole doping, is observed.\cite{plate}  This is exactly what one expects from band theory.  Something quite different is observed in the underdoped regime.  ARPES shows four ``Fermi arcs'' centered at the nodal points near ($\pm\pi/2,\pm\pi/2$).\cite{kmshen}  How does one explain the observation of pieces of Fermi surface which are neither closed orbits nor open orbits intersecting the Brillouin zone boundaries?  One possibility is that there are small hole pockets centered at the nodal points, but due to matrix element effects, only one side of each pocket is  visible in the experiments. Alternatively, there are strong correlation theories which can account for such arcs.\cite{arcs}  Furthermore, the observed arcs are temperature dependent and, at least in some cases, it has been shown that the arcs extrapolate to nodal points at zero temperature.\cite{arpesnodal}  All of these scenarios, Fermi arcs, Fermi nodal points, or small hole pockets in the absence of any long range order which breaks a symmetry, are incompatible with Fermi liquid theory and Luttinger's theorem and do not connect smoothly to the large Fermi surface observed at larger dopings. (Luttinger's theorem says that the area enclosed by the Fermi surface is the same as for non-interacting electrons.)  Furthermore, in the underdoped regime, the superfluid density scales with hole doping (despite band theory predicting a less than half filled electron band) which suggests that this regime is more closely connected to the Mott insulating antiferromagnetic phase at zero doping than it is to the metallic phase of the overdoped regime.  These and other results have led most groups to focus on non-Fermi-liquid descriptions of the pseudogap phase of the cuprates.

Therefore, it came as a surprise when Proust, Taillefer and coworkers\cite{louis} observed quantum oscillations in the longitudinal and Hall resistance of underdoped YBCO, apparently establishing the existence of a well-defined Fermi surface when the superconductivity is suppressed by a magnetic field.  The cross-sectional area of the Fermi surface can be extracted from the period of these oscillations.  Experimental data for the Hall resistivity, exhibiting three clear periods, is shown in Fig.~\ref{ck:fig2}(a).  More recent data shows up to eight periods, leaving little doubt that the period is proportional to the inverse magnetic field.\cite{newdata}  The Fermi surface area extracted from these data is tiny, about 30 times smaller than the Fermi surface area observed in the overdoped regime, and too small to be consistent with hole pockets centered at the nodal points and enclosing p holes per Cu, where p is the hole doping of the sample.  Furthermore, the negative Hall coefficient at low temperatures at this doping of YBCO is taken as evidence that the carriers are electrons, not holes.\cite{el-hole}  However, small electron pockets are incompatible with Luttinger's theorem.  This led Taillefer and coworkers to propose a Fermi surface reconstruction, leading to hole pockets near the nodal points and electron pockets near the zone boundaries as shown in Fig.~\ref{ck:fig2}(b).  This proposal is then compatible with Luttinger's theorem, but raises several other questions.

\def\figsubcap#1{\par\noindent\centering\footnotesize(#1)}
\begin{figure}[b]
\begin{center}
\parbox{2.1in}{\epsfig{figure=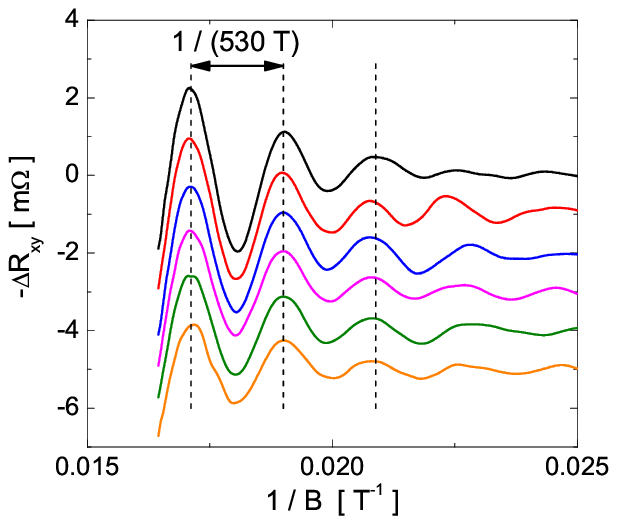,width=3in}
 \figsubcap{a}}
 \hspace{1.8cm}
 \parbox{2.1in}{\epsfig{figure=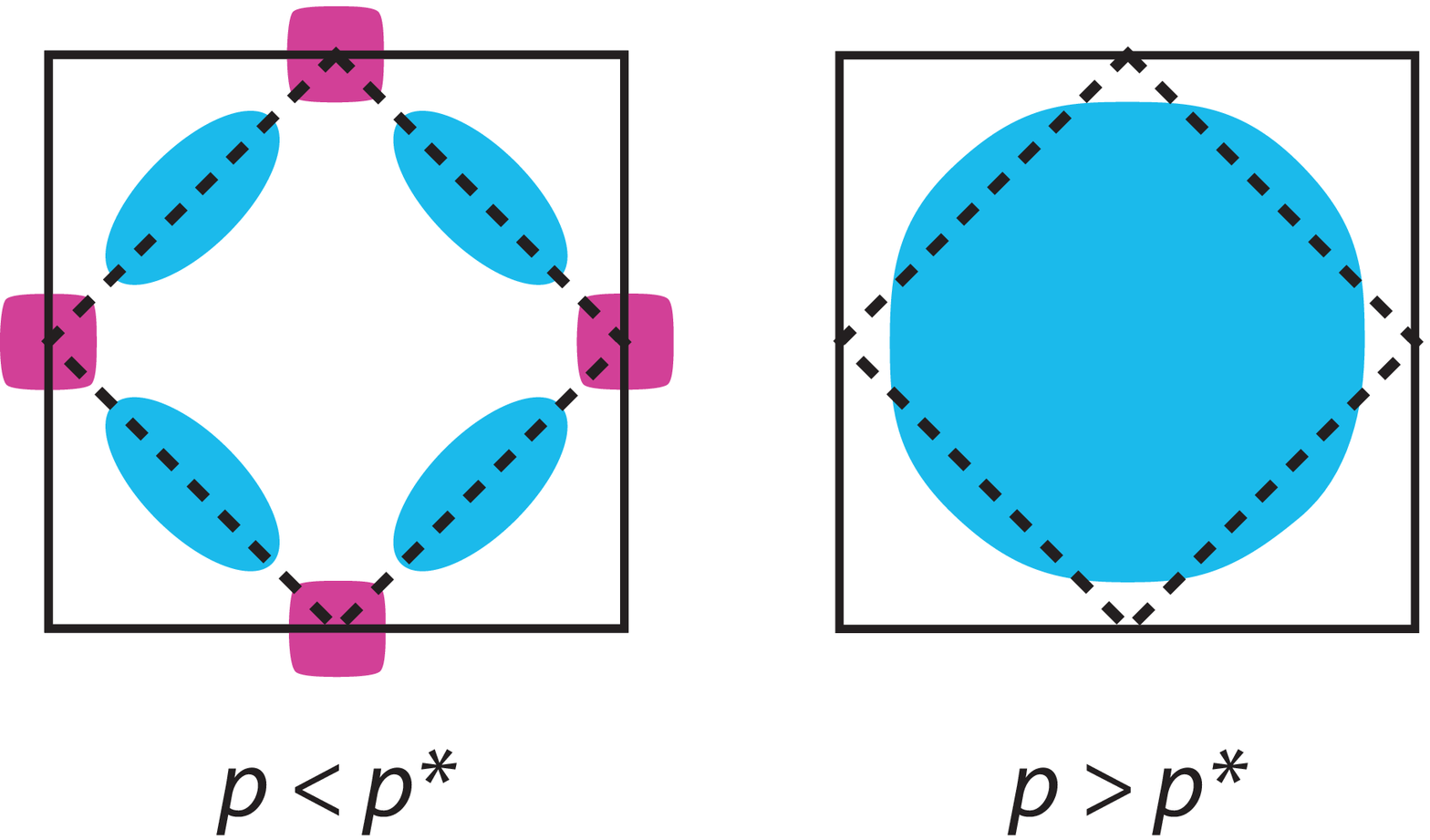,width=2in}
 \figsubcap{b}}
 \caption{Quantum oscillations and Fermi surface, taken from Ref.~\refcite{louis}.
 (a) Oscillatory part of the Hall resistance of underdoped YBCO as a function of inverse field, 1/B, at temperatures ranging from 1.5K (top  curve) to 4.2K (bottom curve). (b) Reconstructed Fermi surface proposed to explain quantum oscillation data in the underdoped (pseudogap) region and the large Fermi surface observed in the overdoped region.}
\label{ck:fig2}
\end{center}
\end{figure}

First, this type of Fermi surface reconstruction is what one would expect in the presence of charge or spin density wave order which introduces a new periodicity.  Possibilities include antiferromagnetism, d-density wave order,\cite{ddens} or stripe order.\cite{stripeorder}  However, no such long-range order has been observed in the pseudogap phase.  It has been suggested that sufficiently slow fluctuations in any of these orders might account for the proposed reconstruction, although this suggestion remains to be quantified and compared in detail to quantum oscillation and other experiments.  A second point which needs to be addressed is the absence of any signature of hole pockets in the quantum oscillation (SdH and dHvA)  measurements, although this could be explained if, because of higher mobility, the electron pockets dominate the signal at low temperatures.  Finally, there is the question of how to reconcile the picture of a reconstructed Fermi surface with ARPES data which sees only Fermi arcs near the nodal points.  In particular, the quantum oscillation data appear inconsistent with the proposal that the Fermi arcs extrapolate to nodal points at zero temperature.\cite{arpesnodal}  However, the observation of Fermi arcs may be compatible with a reconstructed Fermi surface if the rest of the Fermi surface (the other side of the hole pockets, as well as the electron pockets) are obscured by inelastic scattering and matrix element effects.  Another possibility is that the quantum oscillation measurements are probing a different, high magnetic field state, and not the zero field state probed by ARPES.  For example, it has been proposed that antiferromagnetism might be induced by a field and lead to the low oscillation frequency observed.\cite{chen}

At face value, the SdH and dHvA measurements suggest that even the underdoped cuprates might be explained within a Fermi-liquid picture.  However, this suggestion is controversial.  First, the measurements can only be simply explained within a Fermi-liquid picture if there is symmetry breaking order or near-order.   Second, this directly contradicts the suggestion that the pseudogap phase is a nodal liquid and that the Fermi arcs observed by ARPES extrapolate to nodal points at zero temperature. Very recently, Varma\cite{newvarma} has proposed that the quantum oscillation measurements might be compatible with a nodal liquid.

More work is needed to fully understand the implications of observing quantum oscillations in the pseudogap phase and to distinguish between the various possibile proposals for reconciling these data with the ARPES results.     Studies on cuprate materials with different elastic and inelastic scattering rates, as well as different experimental probes, could shed light on reconciling the ARPES and the SdH and dHvA measurements.  For example, Varma\cite{newvarma} suggests infrared absorption measurements to distinquish between a nodal liquid and a reconstructed Fermi surface. Also, the transition implied by Fermi surface reconstruction, whether it is induced by doping or by magnetic field, should show up in other experimental probes.

The nature of the pseudogap phase is still an open question despite more than a decade of intense effort focussed on this one phase. There is convincing evidence for both precursor order (or fluctuations) with a T* which hugs the superconducting dome, and for competing order (or fluctuations) with a T* which cuts through the superconducting dome.  The latter is expected to end in a zero temperature quantum critical point under the superconducting dome.  In fact, both of these phenomena can sometimes be seen in a single experiment.  For example, scanning tunneling measurements see both a ``pairing temperature'' and a pseudogap temperature above \tc.\cite{yazdani}  This and the fact that multiple types of order or quasi-order are observed in at least some materials in some regions of the pseudogap phase, have complicated the identification of the key and universal features of the low-temperature pseudogap phase.

Nevertheless, the existence of a quantum critical point under the superconducting dome, even if precursor effects are also present, would seem to be a key ingredient to understanding high temperature superconductivity, and indeed a variety of proposals exist for the nature of the phases separated by such a quantum critical point.  Most of these suggest a non-Fermi liquid state on the low-doping side, which makes the development of a complete theory of high temperature superconductivity particularly challenging.  Our conventional theoretical formalisms of BCS theory and beyond break down for non-Fermi liquid states and, while our physical understanding of non-Fermi liquid states has deepened considerably and detailed models and calculations exist for highly correlated insulating states, our ability to calculate properties of metallic non-Fermi liquid states, except in one-dimension, is still very limited.  A breakthrough in this area of theoretical physics, might finally allow a complete and predictive theory of high temperature superconductivity.

Given the intense effort and many ideas with strong supporting experimental evidence, it seems likely that the key to the pseudogap phase lies in one of the already existing theories.  Certainly many individuals believe this is the case, but they do not all agree on which theory it is.  As is already clear from just the one class of experiments discussed in detail above, further experiments are likely to confirm or rule out some of the possibilities.  Consequently, this remains a very active area with the hope that new experiments, together with further advances in developing a robust theoretical framework which allows a thorough investigation of metallic non-Fermi liquid states, will resolve open questions in the not too distant future.

\subsection{Iron Pnictide Superconductors}

High temperature superconductivity was discovered in the iron pnictides just last year.  In February 2008, superconductivity at 26K was discovered in LaO$_{1-x}$F$_x$FeAs\cite{hosono}, which quickly rose to 43K in SmO$_{1-x}$F$_x$FeAs,\cite{chen-fe} and 55K in PrO$_{1-x}$F$_x$FeAs.\cite{zhao}  Again, many different materials belong to the class of superconducting iron pnictides.  They fall into two families, referred to as 111 and 122 because of their chemical composition; i.e. LiFeAS and ROFeAs, where R=Ce,Pr,Nd,Sm,... are 111's and (A,K)Fe$_2$As$_2$,  A=Ba,Sr are examples of 122's.  These materials, while containing no Cu, have many similarities to the cuprates and the key question right now is just how similar the pnictides and the cuprates are.  In other words, is the physics of the high temperature superconductivity in the iron pnictides essentially the same as in the cuprates or, are the differences sufficiently important that a new route to high temperature superconductivity has been discovered?  In either case, assuming that electron correlations play a key role in the iron pnictides, as seems most likely, these are extremely interesting materials, and the effort expended per unit time on studying these materials has been even more intense than for the cuprates.  In part, this is because the community has developed many highly relevant tools (both in theory and experiment) from investigations of the cuprates, which can now be quickly redeployed toward the iron pnictides.  For example, in the early days of cuprate research, ARPES was unable to give definitive information, but the precision of ARPES has improved to the point where it is now a central tool for the investigation of high temperature superconductivity.

\begin{figure}[t]
\begin{center}
\psfig{file=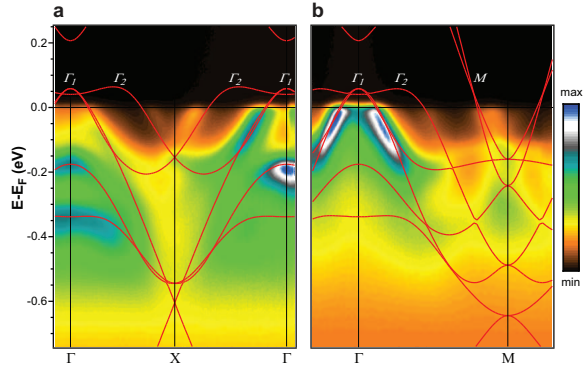,width=3in}
\end{center}
\caption{ARPES data from LaOFeP compared with LDA band structure calculations which have been shifted up by ~0.11eV and renormalized by a factor of 2.2 (red lines).  From Ref.~\refcite{fearpes}.}
\label{ck:iron}
\end{figure}

As mentioned above, the iron pnictides have much in common with the cuprates.  They are both layered materials, with the FeAs layers playing the same role as the CuO$_2$ layers. Both involve d-electrons (from either Fe or Cu) playing a key role; both have antiferromagnetism and superconductivity in close proximity; and both are poor metals which become high temperature superconductors as the temperature is lowered or the doping is increased.  However, there are also differences which may be important.  One key difference is the band structure of the undoped compounds.   The undoped cuprates have one electron per unit cell, so one is starting from a half filled band, which electron correlations turn into a Mott insulator.  In contrast, the undoped iron pnictides have 6 electrons per unit cell which would be a band insulator if the bands did not overlap in energy.  Because the bands do overlap in the pnictides, one is starting from multiple nearly filled or nearly empty bands.  While there is evidence that the band structure is modified, perhaps even significantly, by electronic correlations, the undoped phase remains weakly conducting with five bands crossing the Fermi energy at zero doping.  This band structure, calculated within local density functional theory,\cite{dftbands} agrees reasonably well with what is observed in ARPES experiments for LaOFeP,\cite{fearpes} as shown in Fig.~\ref{ck:iron}.  Recent work suggests that the effects of correlations may be more significant in LaOFeAs.\cite{zxshen}

From the band structure, one would expect electron correlations to play a much smaller role in the iron pnictides.  In the cuprates, at one electron per site, the on-site Coloumb repulsion is extremely important and, in fact, leads to insulating behavior.  In the iron pnictides, each band is almost empty or almost full, so within a band the electrons are far apart and the on-site intraband Coulomb repulsion is not so important.  Interband electron interactions are also reduced because the wave functions are orthogonal.  In a single band model, the Mott insulator transition as a function of onsite Coulomb repulsion, U, occurs roughly at the point where U is equal to the bandwidth.  From the above arguments, in a multi-band model with nearly filled and empty bands, the critical U may be noticeably larger than the average bandwidth. However, there are other signatures, including the fact that the materials are poorer conductors than the band structure would suggest, which have led some to conclude that correlations do, in fact, play a very significant role and that one may be in close proximity to a Mott insulating phase, even though an insulating phase does not appear in the physical phase diagram of these materials.  Furthermore, these materials display commensurate magnetism, which suggests strong correlations.

As mentioned above, antiferromagnetism and superconductivity exist in close proximity, even co-existing in some of the iron pnictide materials.  In contrast to the cuprates, the antiferromagnetism is itinerant, although it is also commensurate and appears at ($\pi$,0).\cite{mag} The observed moment is small, typically less than 0.5$\mu_B$, compared to the 2.5$\mu_B$\ expected from Hund's rule.\cite{mag, moment}  The role of Fermi surface nesting and localized exchange interactions in the observed magnetism is still an open question.  The moment is reduced by more than one would expect simply from quantum fluctuations, and it has been argued that the small moment could arise due to combined effects of spin-orbit, monoclinic distortion and p-d hybridization.\cite{phillips}

The symmetry of the superconducting gap can give important information about electronic correlations and the pairing mechanism.  Given the close proximity of superconductivity and antiferromagnetism in these materials, it is natural to think that antiferromagnetic fluctuations might be driving the superconductivity.  In the cuprates, the strong on-site Coloumb repulsion both stabilizes the antiferromagnetic insulating phase and drives the d-wave symmetry of the superconducting order parameter.\cite{zhang-rice} In general, higher angular momentum pairing is typically a signature of relevant repulsive interactions.

Clearly, there is great interest in determining the symmetry of the superconducting order parameter for the iron pnictides.  To date, there is both evidence for nodes in the gap and for an isotropic (s-wave) gap.  In particular, ARPES measurements have been taken as evidence for an isotropic gap on all five Fermi surfaces in one of the 122 compounds.\cite{ding}  Thermodynamic measurements have seen power law behavior which is taken as evidence for nodes, although one needs to exercise caution in multiband materials.  The superconducting gap can be quite small on some of the bands, making it difficult to distinguish between s-wave and higher angular momentum pairing. In fact, specific heat data was shown to fit a two-gap model well.\cite{nodes}  At this moment in time, the data appears to point toward s-wave in the Fe superconductors, but we do not yet have the definitive measurements that exist in the cuprates, in particular, the phase sensitive measurements.  NMR gives evidence of singlet pairing, again compatible with s-wave pairing.\cite{imai}

Realistic calculations appear to have ruled out a pure phonon mechanism for the iron pnictides.\cite{phonons}  However, Tesanovic has proposed a combined phonon and electronic mechanism, where the phonons provide the attraction within a band and the bands are coupled through electron-electron interactions.\cite{tesanovic}  In this theory, the interband interactions set \tc, much like a Josephson coupling between phonon-mediated superconducting layers would set \tc.  While this theory predicts s-wave gaps, the sign of the gap may differ on different bands.  In principle, one can search for this ``sign-effect'' experimentally.  Finally, there are many purely electronic theories, some which start from an intinerant state with spin fluctuations mediating the superconductivity and others which start with localized moments, a large on-site repulsion and proximity to a Mott insulator, so that much of what we have learned from the cuprates can be applied.

In summary, this field is still very new and is still changing rapidly.  There is currently no consensus on the key question posed here: is high temperature superconductivity in these materials essentially the same as or different from superconductivity in the cuprates.  Directly connected to this question is whether an on-site Coulomb repulsion plays a key role in stabilizing the magnetism and the superconductivity, as it does in the cuprates.  This field is moving more rapidly than the cuprate studies in the early days because we have many more accurate techniques and probes, in theory and in experiment, available to us.  However, one still needs high quality samples for many investigations and creating high quality materials is a mixture of science and art and takes time.  Single crystals have recently become available, but one can expect further advances to be made in removing sources of inhomogeneity and disorder from the crystals.

\subsection{Strontium Ruthenate}

Superconductivity in strontium ruthenate, Sr$_2$RuO$_4$\ was discovered in 1994.\cite{maeno}  The transition temperature, \tc, is low, only 1.5K, but interest in this material stems from the fact that the superconductivity is believed to have a chiral p-wave order which spontaneously breaks time reversal symmetry. Such chiral order would be a solid state analogue of the A phase of He-3 and would also imply a topological order with the potential for exotic physics relevant to quantum computing, as discussed below.  For this reason, much of the effort on Sr$_2$RuO$_4$\ focusses on unambiguously determining the nature of the order parameter.  While there exists strong evidence for chiral p-wave order, some inconsistencies and puzzles remain,  as will be discussed.

Sr$_2$RuO$_4$\ is another quasi-two dimensional material with the same crystal structure as the cuprates.  The electronic action takes place in the RuO$_2$\ layers.  Three bands cross the Fermi energy, and one of these (the $\gamma$\ band composed of $d_{xy}$\ orbitals) is believed to nucleate the superconductivity, with induced superconductivity on the other two bands.\cite{srobands} The transition temperature, \tc, is sensitive to disorder, which immediately suggests that the pairing is likely to be of the unconventional (non-s-wave) type for which scattering around the Fermi surface can average the gap to zero.  Furthermore, early NMR measurements of the Knight shift found that the spin susceptibility was unchanged as the temperature varied through \tc.\cite{nmrsro} This is in contrast to the behavior expected for a conventional s-wave superconductor, where the spin susceptibility falls off rapidly below \tc\ as the spins condense into singlets.  Therefore, the NMR results point to triplet pairing, of which the simplest possibility is a p-wave order parameter, although f-wave has not been ruled out.  About the same time as the NMR results, muon spin resonance (muSR) experiments measured an additional muon spin relaxation which rises from zero at \tc\ and which achieves a maximum value as T approaches zero.\cite{musr}  This extra relaxation was found to correspond to inhomogeneous internal fields with a characteristic strength of a few Gauss.  Since these internal fields are zero above \tc, this experiment points to spontaneous time reversal symmetry breaking in the superconducting state. Recent experiments found the onset of the extra relaxation tracks \tc\ as \tc\ is varied by increasing disorder, reinforcing the interpretation that the time reversal symmetry breaking is directly associated with the superconducting state.\cite{newmusr}

With the experiments pointing toward a triplet, p-wave superconductor with broken time reversal symmetry, the question is which p-wave order parameters are compatible with the symmetry of strontium ruthenate.  There are many allowed p-wave order parameters, as summarized in table IV in Mackenzie and Maeno.\cite{maeno}  In zero magnetic field, one can assume that non-unitary order parameters have higher energy, since they break the symmetry between up and down spins.  Of the unitary p-wave order parameters, there is only one which breaks time reversal symmetry.  It has an isotropic gap around the Fermi surface so it is energetically favorable because of the large condensation energy.

The order parameter for a triplet superconductor must specify the pairing amplitude for each of the three spin states and this can be expressed in terms of a d-vector which contains information about the symmetry of the gap and orientation of the spins:
\begin{equation}
\Delta({\bf k})=i({\bf d}({\bf k})\cdot\vec\sigma)\}\sigma_y
\label{ck:eq1}
\end{equation}
where the components of $\vec\sigma$\ are the Pauli matrices.  For unitary (${\bf d \times d^*}=0$) states, the spin is zero along the direction of ${\bf d}$.  The unitary p-wave state with broken time reversal symmetry corresponds to ${\bf d}=\Delta_0(k_x\pm ik_y)\hat{\bf z}$, which has a  chirality given by the $\pm$\ sign.  The two chiralities are degenerate, so there is the possibility of domain structures. Due to spin-orbit coupling in strontium ruthenate, the d-vector is oriented along the c-axis (chosen to be the z-axis here) so the spins are in the $S_z=0$ state.  This also corresponds to equal spin pairing ($\uparrow\uparrow$\ and $\downarrow\downarrow$) in the ab (or xy) plane.  In this state, each Cooper pair carries angular momentum plus or minus one, depending on the chirality, along the z-axis.  The BCS wave function carries a total angular momentum of $N\hbar/2$, where $N$\ is the total number of electrons.

The BCS state described by this chiral p-wave order parameter is a two-dimensional analog of the A phase of He-3,\cite{he3} and is also closely related to the Moore-Read state proposed for a quantum Hall system at 5/2 filling.\cite{moore-read}   As shown by Moore and Read, the 5/2 state has a topological order and supports Majorana zero modes at the edges and at vortex cores.  Majorana fermions are their own antiparticle (i.e., $\gamma^\dagger = \gamma$, where $\gamma^\dagger$\ creates a Majorana fermion) and two Majorana fermions are required to create an ordinary fermion, such as an electron.  Much exotic physics, including non-Abelian statistics follows from the fact that this state supports Majorana fermions.

Even if strontium ruthenate does support a chiral p-wave state, the exotic physics is not immediately accessible because the direct correspondence is between the Moore-Read 5/2 state and a {\em spinless} (or, equivalently, spin polarized) chiral p-wave superconducting state.
The equal spin pairing state appropriate for strontium ruthenate is equivalent to two copies (spin up and spin down) of the Moore-Read state and, consequently, supports two Majorana zero modes at the edges and at vortex cores and much of the exotic physics is lost.  However, if the d-vector can be rotated into the ab-plane, and is free to rotate in that plane, the exotic physics predicted in the Moore-Read state becomes accessible.

A d-vector which is free to rotate in the ab-plane corresponds to pairing in only a single spin channel ($\uparrow\uparrow$\ or $\downarrow\downarrow$) which suggests it might be stabilized by an external magnetic field.  In fact, recent NMR experiments have been interpreted as evidence for such a state.  Earlier NMR experiments were done with a magnetic field in the ab-plane and saw no suppression of the spin susceptibility below \tc, as one would expect for a triplet state with equal spin pairing in the ab-plane.  However, more recent NMR experiments with the magnetic field along the c-axis also found no suppression of the spin susceptibility below \tc.\cite{newnmr}  This is not compatible with a $S_c=0$ state and has been taken as evidence that modest fields (less than 500G) are sufficient to rotate the d-vector into the plane.  In He-3, which is isotropic, it is known that magnetic fields rotate the d-vector perpendicular to the field.  The spin-orbit coupling in strontium ruthenate is sufficiently strong that it is surprising such low fields would reorient the spins.  On the other hand, one needs to compare the energies of the different states in the presence of a field.  It has been argued that there is, in fact, an energetically competitive state with the d-vector in the plane.\cite{sigristother}  However, this state is non-chiral and, consequently, would not support the exotic physics of the Moore-Read state.  Currently, it is an open question as to what state is stabilized in a c-axis field.  Nevertheless, theorists have explored the possibility of exotic physics if a chiral p-wave state with a d-vector in the ab-plane is stabilized, so let me briefly review some of the highlights of these explorations.

If the d-vector lies in the ab-plane and is free to rotate, the system can support half-quantum vortices.  The wave function acquires a phase of $\pi$\ if the structure of the vortex is such that the d-vector winds around the vortex core.  Therefore, the orbital part of the wave function also only needs to acquire a phase of $\pi$, rather than the usual 2$\pi$ associated with a vortex, for the entire wave function to be single valued.  This corresponds to the Bohm-Aharonov phase of a Cooper pair circling half of the usual superconducting flux quantum, or $hc/4e$.  Whether such a half-quantum vortex has a lower or higher energy than the regular vortex, depends on microscopics, and there have been proposals for stabilizing such vortices.\cite{kim}  One can show that the half-quantum vortex supports a single Majorana zero mode bound at the core.\cite{core,readgreen}  (This is in contrast to the usual vortex which has two zero modes in the core.)  Furthermore, these half-quantum vortices obey non-Abelian statistics when one vortex is moved around another such vortex.\cite{readgreen}  Non-Abelian statistics is exactly what is required in quantum computing, as the non-trivial winding connects distinct, but degenerate ground states with topological stability.\cite{qucomp}  Of course, even if strontium ruthenate does support exotic vortices, one needs to carefully consider the role of the third dimension as one would expect the Majorana fermions to form a band along the c-axis, which will complicate their role in quantum computing.

Having presented evidence for chiral p-wave superconductivity and discussed some of the possible exotic physics which could arise from this state, I now want to turn to the more recent experiments which have provided both further compelling evidence for chiral p-wave order, as well as results which suggest otherwise.  In particular, I will focus on the polar Kerr effect and the search for spontaneous edge currents.

In the polar Kerr effect, linearly polarized light is normally reflected from the sample surface as elliptically polarized light with a rotation of the polarization axis being the Kerr angle.  One observes a non-zero Kerr angle if either left or right circularly polarized light is preferentially absorbed by the sample, as would be the case in a ferromagnet or a chiral p-wave superconductor.  Kapitulnik's group observed a non-zero Kerr angle grow up as strontium ruthenate was cooled below \tc.\cite{kapitulnik}  The Kerr angle rose from zero at \tc\ to a maximum of 60 nrads at the lowest temperatures.  The sign of the Kerr angle, but not the magnitude, was affected by cooling in fields up to 100G.  These data are qualitatively as expected for a chiral p-wave superconductor with a domain size larger than the beam size of incident light.  In some runs a reduced Kerr angle was observed, which suggests the domains are not too much larger than the beam size which is about 25 to 50 microns across.

In a clean chiral p-wave superconductor the idealized Kerr angle is strictly zero from translational symmetry.\cite{readgreen} However, since the beam size is finite, one is not probing the system at strictly zero wave vector, and, in fact, a clean chiral p-wave superconductor displays interesting and nontrivial behavior at finite wave vector.\cite{roy,yakovenko}  However, the beam is large enough in Kapitulnik's experiment that these effects should be negligible.  Recently, Goryo showed that the lowest order impurity induced contribution to the Kerr angle comes from so-called skew-scattering diagrams, which contribute in order $n_iU^3$, rather than the usual $n_iU^2$ term, where $n_i$ is the density of impurities and $U$ is related to the strength of the impurity potential. \cite{goryo} Estimates of the Kerr angle from this impurity scattering model are smaller than, but comparable to, the observed value, if one takes somewhat optimistically large estimates for the density and strength of impurity scattering.  Therefore, this seems like a possible, although somewhat marginal, explanation of the experiments.  This theory could be tested by further experiments, since it predicts an unusual $\omega^{-4}$ frequency dependence for the Kerr angle.  Furthermore, one could try increasing the amount of disorder, while still maintaining superconductivity (at a reduced \tc) to test this interpretation.  Nevertheless, while some questions remain, the Kerr effect is a very direct probe of time reversal symmetry breaking and chirality and these experiments significantly strengthen the case for chiral p-wave superconductivity.  As a final point, it is interesting that Goryo's theory only gives a non-zero result for p-wave and would give zero for a chiral f-wave superconductor.\cite{goryo}

\begin{figure}
\begin{center}
\psfig{file=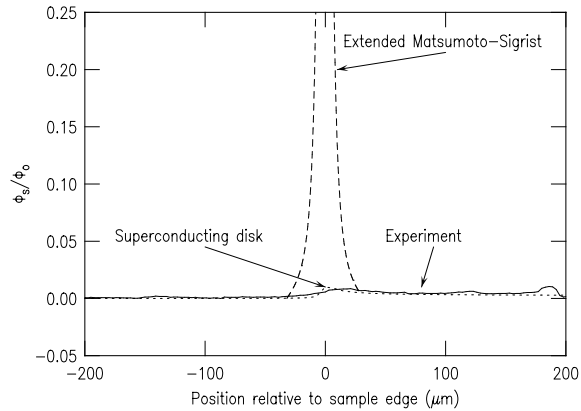,width=3in}
\end{center}
\caption{SQUID scan across the edge of an ab face of a Sr$_2$RuO$_4$\  crystal (solid line). The dotted line is the prediction for an s-wave superconducting disk in a uniform residual Þeld of 3 nT. The dashed line is the prediction for a single domain $p_x+ip_y$\ superconductor, following the theory Matsumoto and Sigrist,\cite{matsumoto} but modifed for a finite sample. The peak value of the dashed line is 1 (off-scale). From Kirtley {\it et al.}, Ref.~\refcite{kirtley}. }
\label{ck:squid}
\end{figure}

Another direct test for chiral p-wave order is to search for spontaneous supercurrents flowing at the sample edges and/or at domain walls.\cite{yakedge}  In fact, the early muSR experiments are interpreted as evidence for supercurrents at domains walls in the bulk, since the magnetic field inside a single-domain chiral p-wave superconductor vanishes (except at defects which suppress the superconductivity, such as at impurity sites).  The topological nature of the state, requires special edge modes at zero energy, but in addition, a chiral p-wave state supports a band of edge modes which carry a spontaneous supercurrent related to the total angular momentum of the state.\cite{stone}  This supercurrent is localized roughly within a coherence length of the surface and is screened by an equal and opposite current within roughly the coherence length plus the penetration depth.  Consequently, in the absence of domains, the field is strictly zero in the bulk, but there is a net magnetization or field localized at the surface.  Similar currents flow at domain wall boundaries.\cite{matsumoto}  One should be able to detect the fields associated with these currents at the edges or from domain walls intersecting the surface, using scanning SQUID microscopy or a scanning Hall probe.  Both techniques have been employed on strontium ruthenate, and no evidence of fields at the surface were observed.\cite{moler,kirtley}  Fig.~\ref{ck:squid}, for example, shows the experimentally observed flux as one scans across the sample compared to the flux expected for a somewhat idealized chiral p-wave superconductor.  The expected flux is about two orders of magnitude larger than the experimental noise limit.  In fact, the experimental data can be well modeled by an s-wave superconductor screening a residual external field of 3 nT.

These null results are quite surprising and difficult to reconcile with chiral p-wave order.  Very small domains (at the surface) could explain the null results because of the finite size of the pickup loops (8 microns for the SQUID and 0.5 microns for the Hall bar).  Such small domains, roughly a micron or smaller, would be incompatible with the measured Kerr angle. Furthermore, domain walls cost energy and are expected to be present at low temperatures only due to pinning effects.  Rough or pairbreaking surfaces, as well as other modifications to the theory, can reduce the expected signal, but no plausible explanation has been found which both reduces the signal to below the experimental sensitivity and leaves the interpretation of the positive experiments, such as the Kerr effect and muSR measurements,  intact.\cite{kallinberlinsky,ashby}  At the moment this is a puzzle, and it is interesting to note that the same puzzle persists for the A phase of He-3.  The symmetry of the A phase has been established without a doubt, due to high precision measurements of the various collective modes, for example.\cite{he3}  However, the mass supercurrents expected at the surface have never been observed, although in He-3, the d-vector is free to rotate and may do so near the boundary, which could suppress these currents.

The absence of observed currents in He-3 led Leggett to suggest an alternative to the BCS wave function.\cite{leggett}  While the BCS wave function carries an angular momentum of $N\hbar/2$, for Leggett's wave function this is reduced by a factor of $(\Delta/E_F)^2$.  This would certainly make the supercurrents unobservable, although it would also eliminate the explanation of the muSR results in terms of fields associated with domain walls.  However, I believe it would leave the Kerr effect interpretation intact.  In any case, I think it still remains to be understood whether the weak-coupling limit of a chiral p-wave superconductor is  described by the BCS wave function or an alternative, such as Leggett's wave function.  For the case of s-wave, the two wave functions are identical.

In summary, there is compelling evidence pointing toward chiral p-wave order in the superconducting state of strontium ruthenate.  In addition to the muSR and Kerr effect results discussed above, there are also several tunneling results which point toward chiral p-wave order.\cite{vanhar,liu}  The absence of observed edge currents remains a puzzle which is difficult to reconcile with chiral p-wave order.  Furthermore, if one looks closely at the details of the various experiments, one finds that all the experiments are relying on certain assumptions about domain sizes.  Some experiments (such as the Kerr effect) require the domain size to be sufficiently large to interpret the experiment as evidence for chiral p-wave order, whereas other experiments require the domains to be sufficiently small (e.g. muSR).  Consequently, the experiments are not as consistent with each other as one might first assume, and ideally one would like to be able to probe the domain walls directly, if they do exist.  Certainly more work needs to be done to unambiguously determine the symmetry of the superconducting order.  The striking observations of time reversal symmetry breaking, together with theories which point toward exotic physics and potential applications to quantum computing, provide significant motivation for further studies on this material.

\section{Frustrated Magnets}

The great challenge driving the search for new frustrated magnetic materials is to discover a material which supports a two or three dimensional quantum spin liquid.  Spin liquids are ubiquitous in one-dimensional magnetic systems since quantum fluctuations prevent order.  So far, spin liquids in higher dimensions have remain elusive in real materials despite an aggressive search over the last two decades.   Theoretical models exist in higher dimensions, both for gapped spin liquids, which exhibit topological order, and gapless spin liquids which may have a spinon Fermi surface.\cite{sachdev-solvay}  This field, including the relevant experiments, is reviewed by Subir Sachdev, so I will keep my discussion brief and just highlight a few points specific to the real materials currently under investigation.   The best experimental candidates are typically spin 1/2 systems, for which quantum effects are maximized, either with geometric frustration and macroscopic classical degeneracy, or with proximity to a metal-insulator transition so that fluctuations, in particular ring exchanges, are important. Examples of the first kind include Herbertsmithite (ZnCu$_3$(OH)$_6$Cl$_2$), Volborthite (Cu$_3$V$_2$O$_7$(OH)$_2$H$_2$O), and Vesignieite (BaCu$_3$V$_2$O$_8$(OH)$_2$), which all correspond to spin 1/2 on a Kagome lattice.  Examples of the second kind include the organics $\kappa$-(BEDT-TTF)$_{2}$Cu$_{2}$(CN)$_{3}$ and EtMe$_3$Sb[Pd(dmit)$_2$]$_2$, with spin 1/2 on a triangular lattice, and close to a Mott insulator transition.  All of these systems, as well as many more examples, exhibit no conventional magnetic order down to the lowest temperatures studied, often several orders of magnitude below the Curie-Weiss temperature which is inferred from high temperature susceptibility measurements.

While theoretical studies suggest either frustration and degeneracy or proximity to a Mott transition as conditions conducive to spin liquid behavior, real materials typically present challenges which complicate the search for a spin liquid.  In particular, in materials which rely on geometric frustration and degeneracy, one seldom achieves the perfect or near perfect lattice structure.  Typically the lattices are either distorted from the ideal lattice configuration or there is intrinsic disorder which is difficult (perhaps even impossible) to eliminate or both of these effects occur.  If one runs through the long list of materials based on quantum spins on Kagome or pyrochlore lattices, it seems that one can obtain clean materials with distorted geometry or disordered materials with ideal geometry, but, at least to date, not clean materials with ideal geometry.  In other words, it seems that nature at least partially lifts the macroscopic degeneracy through either spontaneous distortion or disorder, rather than through the quantum fluctuations which would lead to a uniform spin liquid.  For example, in Herbertsmithite, the Cu atoms form Kagome layers, but there is noticeable exchange of Zn atoms, which sit between the layers, and the Cu atoms.  This is seen in NMR where one observes two different O sites, depending on whether one of the neighboring Cu is replaced by Zn or not.\cite{olariu}  Such disorder reduces the frustration, lifts the classical degeneracy and can affect many spins.  At best, this can make the identification of the spin liquid state difficult and at worst, it can lead to a more conventional, but disordered, state. Nevertheless, materials discoveries often surprise us, and one may yet discover a material with a more ideal Kagome structure.

The second route to a spin liquid, proximity to a Mott insulating transition, does not rely on an underlying macroscopic classical degeneracy.  The examples of organic compounds, mentioned above, have spins on a triangular lattice.  Here, the spin 1/2's reside on large molecules, and disorder may also present problems but it does not play the role of partially lifting a necessary condition for the route to a spin liquid.  For this reason, this second class of materials, which includes $\kappa$-(BEDT-TTF)$_{2}$Cu$_{2}$(CN)$_{3}$, may be particularly promising materials for finding a spin liquid in two or three dimensions.

\section{Topological Insulators}

The discovery of the quantum Hall effect in 1979 ultimately opened up a new field of study which connects all the topics discussed here, namely, the field of topological order.  The integer quantum Hall state is an example of a topological state which has no conventional broken symmetry and is not described by a local order parameter, but, rather, is characterized by a topological invariant, the first Chern number.  Recent advances related to the quantum Hall effect are reviewed by Ady Stern.\cite{stern-solvay}  Here I discuss recent experimental discoveries of the quantum spin Hall effect and topological insulators.

The quantum spin Hall (QSH) state was first predicted in 2005,\cite{kanemele, bernevig} and discovered experimentally in 2007.\cite{koenig}  It is a topological state, closely related to the integer quantum Hall state but it does not require an external magnetic field and, in fact, arises in systems with time reversal symmetry.  It occurs in two-dimensional systems with a non-trivial band structure arising from strong spin-orbit interactions, such that the system is an insulator in the bulk but supports topologically protected edge states.  These edge states are analogous to the chiral edge states in the integer quantum Hall effect, and the QSH state can be thought of as two copies of quantum Hall states, one for each spin component, which move in opposite directions.

The QSH state, a new state of matter, was observed in HgTe quantum wells surrounded by CdTe,\cite{koenig} as predicted by theory. The observed conductance is independent of the width of the well, as expected for a conductance due to edge states only.  Furthermore, its magnitude at low temperatures is the expected quantized value of $2e^2/h$, provided the sample is not too long (along the direction of the edge currents).  In long samples, the conductance is suppressed, as the QSH currents are only protected by time-reversal symmetry.  Molenkamp and coworkers\cite{koenig} also verified that the QSH effect was destroyed by applying a magnetic field.

The QSH effect is not restricted to two dimensions and topological insulators also exist in three dimensions.\cite{3dtop}  Again, these are systems with a non-trivial band structure due to strong spin orbit interactions and which support conducting, topological edge states.  These insulators are distinguished from ordinary insulators by a Z$_2$ quantum number which takes the value $\nu=0$\ for ordinary insulators and $\nu=1$\ for topological insulators.

\begin{figure}[t]
\begin{center}
\psfig{file=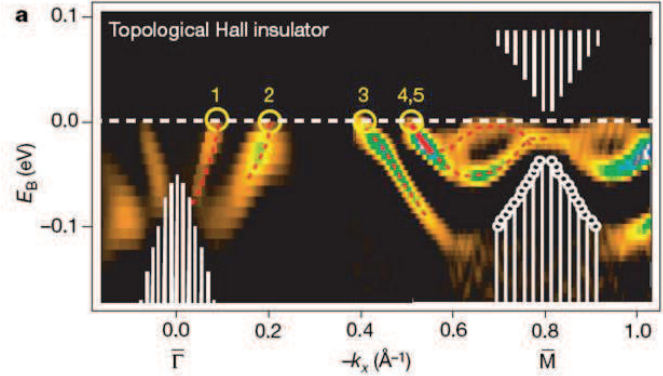,width=3.5in}
\end{center}
\caption{ARPES data showing the surface band dispersion of Bi$_{0.9}$Sb$_{0.1}$\ along $\Gamma$--M.  The Fermi crossings of the surface state are denoted by yellow circles, with the band near $-k_x\approx 0.5 \AA^{-1}$\
counted twice owing to double degeneracy.  From Ref.~\refcite{hsieh}.}
\label{ck:topins}
\end{figure}

The key difference between ordinary and topological insulators can be understood by focussing on the properties of the edge states.  While an ordinary band insulator can support edge (or surface) states, these edge states are not topologically protected, any crossings (degeneracies at the same point in k-space) typically occur at general points in the Brillouin zone, and perturbations will open up a gap at these crossings.  In topological insulators, the special edge states occur at symmetry points (actually at Kramers degeneracy points, such as $\Gamma$ and M).  Time reversal symmetry requires that these surface states come in Kramers pairs and protects them against perturbations.  One can have a single Kramers pair at these symmetry points.  It follows that topological insulators ($\nu=1$) have an odd number of surface states crossing the Fermi energy between the points $\Gamma$ and M, say, whereas this number must be even for a conventional band insulator.  This has been observed in Bi$_{1-x}$Sb$_x$\ in a beautiful set of experiments.\cite{hsieh}  High resolution ARPES measurements found 5 surface states crossing the Fermi between the $\Gamma$\ and M points. These data are shown in Fig.~\ref{ck:topins}. Care has been taken to identify the surface bands, accounting for multiple bands which are close by in energy, by observing the splittings at other points in k-space, and these measurements provide compelling evidence of a three-dimensional topological insulator. More recently, spin-ARPES was used to probe the spin degrees of freedom and confirm the chirality of the surface states.\cite{hsieh2}

In striking contrast to the field of novel superconductors discussed above, one point that stands out in the field of topological insulators is the detailed, predictive power of theory.  Theorists predicted specific materials to be candidates for topological order, which were then experimentally verified a short time later.  This, of course, is because the novel physics of topological insulators occurs at the non-interacting or one-electron level.  In fact, in the above discussions and in most of the theoretical work, electron-electron interactions are ignored and assumed to be weak.  Much less is known about potential strongly correlated topological insulators and whether there are spin analogues to the fractional quantum Hall effect, for example.  This is currently an active field of study, as is the search for new physics in the non or weakly interacting topological insulators discussed here.  There already are additional theoretical proposals, such as the search for an emergent magnetic monopole induced by external electric field,\cite{zhang} which are awaiting experimental discovery.

\section{Conclusions}

Real materials support new topological states connected to (but distinct from) the quantum Hall effect, such as two and three dimensional quantum spin Hall and topological insulating systems.  In addition, strontium ruthenate may support a chiral p-wave state which is also connected to a topological quantum Hall state.  This illustrates the enormous influence the quantum Hall effect has had on condensed matter physics and the field of quantum and topological order.  It will certainly be interesting to explore the possibility of fractionalization in these topological states, to see if the analogies with quantum Hall physics go even deeper.\cite{sczhang}

On the other hand, quantum ordered or topological states motivated by studies of the high temperature superconducting cuprates (rather than by quantum Hall studies) remain elusive in real materials in dimensions higher than one, despite intense efforts in discovering, creating and improving frustrated magnetic materials.  There has been enormous progress in understanding the theory of quantum spin liquids, both gapped, topological spin liquids as well as gapless, quantum ordered spin liquids.  Many frustrated magnetic materials exhibit correlated spin states at low temperatures with no magnetic order, often down to temperatures which are less than $10^{-4}$ of the Curie-Weiss temperature.  However, while candidates for spin liquids exist, noteably herbertsmithite and the organic $\kappa$-(BEDT-TTF)$_{2}$Cu$_{2}$(CN)$_{3}$, a smoking gun experiment for spin liquid order remains elusive and often intrinsic disorder, lattice distortion or anisotropic interactions play a key role in differentiating the real materials from the theoretical models.

Superconductivity remains a fascinating and active area of research.  Superconductivity shares a property with the other subjects discussed here, quantum spin liquids and topological insulators, in that even ordinary BCS superconductivity is a type of topological order.\cite{sondhi} However, more recently, we have seen that it may be able to support further topological order, such as chiral p-wave order in strontium ruthenate. The cuprates gives us an example of a superconductor with strong repulsive interactions playing a key role. In addition to providing us with the highest superconducting transition temperatures known to date, the cuprates also exhibit the intriguing but puzzling pseudogap phase and appear to support a robust quantum critical point which may be connected to much of the anomalous observed behavior. The discovery of a new class of high temperature superconductors has generated renewed interest, but the question remains  whether these new iron-based superconductors will provide new insights into the phenomenon of high temperature superconductivity or whether they will instead generate a new set of puzzles of their own.

\section*{Acknowledgments}

I would like to thank John Berlinsky, John Kirtley, Steve Kivelson, Sung-Sik Lee, Kathryn Moler, Louis Taillefer, Tom Timusk, and Shoucheng Zhang, for many useful discussions and Louis Taillefer, Z.X. Shen and Zahid Hasan for permission to use their figures. Support from the Canadian Institute for Advanced Research, Canada Research Chairs program and Natural Sciences and Engineering Research Council of Canada is gratefully acknowledged.


\begin{thebibliography}{9}
\bibitem{pseudogs} Despite considerable effort to make high quality crystals at dopings near where the superconducting \tc\ vanishes on the underdoped side, intrinsic disorder appears to be unavoidable and complicates direct experimental studies of the pseudogap phase at the lowest temperatures.
\end{thebibliography}
\end{document}